\documentclass{article}
\usepackage{spconf,amsmath,graphicx}
\usepackage{cite}
\usepackage{amssymb}
\usepackage{multirow}
\usepackage{booktabs}
\usepackage{color}
\usepackage{caption}
\usepackage{url}

\title{Rep2wav: Noise Robust text-to-speech Using self-supervised representations}
\name{Qiushi Zhu$^{1,2}$, Yu Gu$^1$, Rilin Chen$^1$, Chao Weng$^1$, Yuchen Hu$^{3}$, Lirong Dai$^2$, Jie Zhang$^{2}$ 
	\thanks{ This work was done at Tencent AI LAB as an internship by Q. Zhu.}}
\address{ $^1$Tencent AI LAB\\
	$^2$NERC-SLIP, University of Science and Technology of China (USTC), Hefei, China\\ 
 $^3$Nanyang Technological University, Singapore\\ 
}

\begin{document}
\ninept
\maketitle
\begin{abstract}
Benefiting from the development of deep learning, text-to-speech (TTS) techniques using clean speech have achieved significant performance improvements. 
The data collected from real scenes often contains noise and generally needs to be denoised by speech enhancement models.	
Noise-robust TTS models are often trained using the enhanced speech, which thus suffer from speech distortion and background noise that affect the quality of the synthesized speech. 
Meanwhile, it was shown that self-supervised pre-trained models exhibit excellent noise robustness on many speech tasks, implying that the learned representation has a better tolerance for noise perturbations. In this work, we therefore explore pre-trained models to improve the noise robustness of TTS models. 
Based on HiFi-GAN, we first propose a representation-to-waveform vocoder, which aims to learn to map the representation of pre-trained models to the waveform. 
We then propose a text-to-representation FastSpeech2 model, which aims to learn to map text to pre-trained model representations. 
Experimental results on the LJSpeech and LibriTTS datasets show that our method outperforms those using speech enhancement methods in both subjective and objective metrics.
Audio samples are available	at: \url{https://zqs01.github.io/rep2wav}.	
		
\end{abstract}
\section{Introduction}
\label{sec:intro}
Text-to-speech (TTS)~\cite{klatt1987review,6542729,gu2021bytesing} aims to synthesize natural and intelligible speech from input text.
Thanks to advanced deep learning techniques, neural network-based TTS models can synthesize high-quality speech when trained using clean speech data.
However, collecting clean speech data requires quiet environments and high-quality recording equipments, e.g., professional audio studios, resulting in high data collection costs.
Meanwhile, noisy speech data is very easy to collect and is available in large amounts.
If these noisy data can be used for building high-fidelity TTS models, the cost of data collection will be largely reduced and the trained TTS model becomes more applicable.
How to train TTS models using noisy data is therefore the focus of this work.

There are many approaches devoted to training TTS models using noisy speech, where the majority uses speech enhancement models for data denoising and then trains TTS models using the enhanced speech.
For example, pre-trained speech enhancement models were used in~\cite{Valent,valentini2016investigating,8343873} for denoising, followed by training  TTS models with the enhanced speech.
In~\cite{Valent,valentini2016investigating}, recurrent neural networks  were trained using parallel noisy and clean speech and Hidden Markov Model based acoustic models were then trained with the enhanced features.
Although this scheme performs well in simple noise situations, the enhanced speech is susceptible to speech distortion and unseen noise, which can harm the training of the TTS model.
To avoid the use of speech enhancement models, it was suggested in~\cite{dai2020noise,hsu2018hierarchical,8683561,9413934,saeki22_interspeech,yang2022norespeech} to train TTS models directly using noisy data. In~\cite{dai2020noise} an end-to-end TTS model was proposed by using speaker embedding and noise representation as conditional inputs to model speaker and noise information separately.
In~\cite{hsu2018hierarchical,8683561}, the noise representation was taken as input and  the background noise was then removed from the speech via representation decoupling.
As in~\cite{hsu2018hierarchical,8683561} the noise embeddings  are sentence-level vectors with coarse granularity, which might not be suitable for complex noise scenarios, DenoiSpeech was proposed in~\cite{9413934}, which considers fine-grained frame-level noise modeling to handle real-world noisy speech.
DRSpeech was proposed in~\cite{saeki22_interspeech}, which jointly represents time-variant additive noises with a frame-level encoder and an utterance-level encoder.
Although these methods demonstrate a good noise-robust performance, most exploit mel-spectrogram features.

Self-supervised pre-trained models have shown an excellent performance and strong noise robustness on many speech tasks.
In the field of automatic speech recognition (ASR), the self-supervised pre-trained models Wav2vec2.0~\cite{baevski2020wav2vec}, HuBERT~\cite{hsu2021hubert}, Data2vec~\cite{baevski2022data2vec} and WavLM~\cite{9814838} were proposed recently, and in~\cite{pasad2021layer} using the pre-trained models to learn different levels of information at different layers was analyzed.
ASR models such as the problem-agnostic speech encoder (PASE+)~\cite{ravanelli2020multi}, Wav2vec-switch~\cite{wang2022wav2vec} and enhanced wav2vec2.0~\cite{9747379} exhibit an excellent noise robustness in noisy environments.
Based on these methods, in~\cite{chang22g_interspeech,10122599} the combination with speech enhancement models was revealed to further improve the ASR accuracy in noisy scenes.
The enhanced speech is fed into the pre-trained model to reduce the impact of speech distortion, which somehow support the fact that the pre-trained representation has a strong ability to resist speech perturbation.
Similarly, in the field of speech synthesis, TTS models trained using enhanced speech also suffer from the issues of speech distortion and noise.
Whether the above problems can be mitigated using pre-trained models and whether higher quality speech can be synthesized is unexplored.
{\it It is thus questionable whether higher-quality speech can be synthesized by leverging pre-trained models in practice.}
In fact, there are some related studies, e.g., ~\cite{siuzdak22_interspeech,du22b_interspeech,lee2022hierspeech} on training TTS models using representations, however the noise robustness of representation-based TTS models is not yet explored.

\begin{figure*}[!t]
    \centering
    \includegraphics[width=0.9\textwidth]{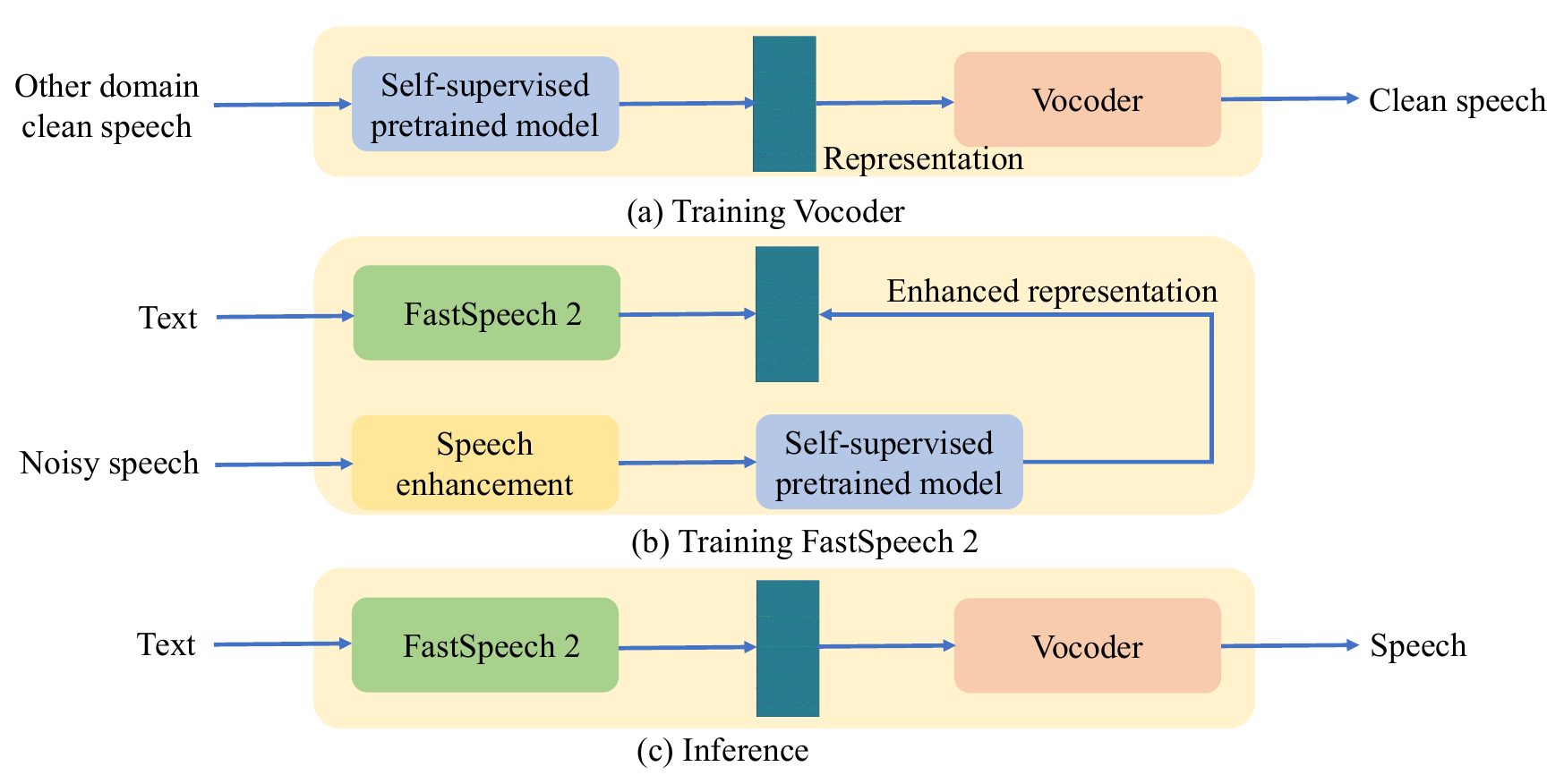}
    \caption{The proposed TTS paradigm: (a) representation-to-waveform vocoder, (b) text-to-representation FastSpeech 2, and (c) inference.}
    \label{fig:figure1}
\end{figure*}

In order to verify that the representation feature has a better ability to resist speech perturbation than the conventional mel-spectrogram for speech synthesis and to improve the quality of the synthesized speech, in this paper we propose a noise-robust TTS model based on the representation.
Based on HiFi-GAN~\cite{kong2020hifi}, we first propose a representation-to-waveform vocoder, which aims to learn to map the representation of the pre-trained model to the waveform. 
Based on FastSpeech 2~\cite{ren2020fastspeech}, we then propose a text-to-representation FastSpeech 2 model, which aims to project text to pre-trained model representations. 
Results show that 1) the higher layer of representation is used, the more contextual information is included in the representation and the better the noise robustness of the model, but meanwhile more speaker information is lost;
2) The use of different layer-averaged representations can balance both the noise robustness of the model and the speaker information;
3) The performance of the TTS model based on representation is better than that of the TTS model based on the mel-spectrogram in terms of both objective and subjective evaluation metrics.

\section{Methodology}
\label{sec:method}
In this section, we introduce the components of our noise-robust TTS model, including the representation-to-waveform vocoder and the text-to-representation FastSpeech 2 model.

\subsection{Representation-to-waveform: Vocoder}
The representation-to-waveform vocoder is based on HiFi-GAN~\cite{kong2020hifi}, which mainly contains a generator and two discriminators, i.e., multi-scale and multi-period discriminators, and both the generator and the discriminator use multi-layer convolutional networks.
The generator takes the representations of different layers of the pre-trained model as input and then upsamples them by multi-layer transpose convolutions until the length of the output sequence matches the temporal resolution of the original waveform.
The discriminator is used to discriminate the signal patterns of different periods in the speech signal.
For the detailed network structure of generators and discriminators, please refer to~\cite{kong2020hifi}.
The procedure for training the vocoder is shown in Fig.~\ref{fig:figure1}(a).
To ensure that the data for training the vocoder is universal, we select publicly available multi-speaker clean speech datasets from other domains.
The clean speech $x$ is fed into the pre-trained model to extract the output representation $c$ from different layers, and then the representation $c$ is fed into the vocoder to reconstruct the clean speech waveform.
Given the generator $G$ and discriminator $D$, the total generator loss function $\mathcal{L}_{G}$ and the discriminator loss function $\mathcal{L}_{D}$ for training the vocoder can be respectively formulated as
\begin{equation}
\mathcal{L}_G=\mathcal{L}_{adv}(G;D)+\alpha \mathcal{L}_{fm}(G;D) + \beta \mathcal{L}_{mel}(G),
\label{eq1}
\end{equation}
\begin{equation}
\mathcal{L}_D=L_{adv}(D;G),
\label{eq2}
\end{equation}
where the generative loss $\mathcal{L}_{adv}(G;D)$ and discriminative loss $\mathcal{L}_{adv}(D;G)$ are respectively given by
\begin{equation}
\mathcal{L}_{adv}(D;G)=\mathbb{E}_{(x,c)}\left[(D(x)-1)^2 + (D(G(c)))^2\right],
\label{eq3}
\end{equation}
\begin{equation}
\mathcal{L}_{adv}(G;D)=\mathbb{E}_{(c)}\left[(D(G(c))-1)^2\right].
\label{eq4}
\end{equation}
The feature matching loss $\mathcal{L}_{fm}(G;D)$ and the mel-spectrogram loss $\mathcal{L}_{Mel}(G)$ in (\ref{eq1}) keep the same as~\cite{kong2020hifi}. $\alpha$ and $\beta$ are hyperparameters.

\begin{table*}[!t]
\caption{The SNR of the synthesized speech by the vocoder when the mel-spectrogram and representation of the 5 dB noisy speech are input, where the vocoder is trained using clean speech.}
\label{tab:table1}
\centering
\begin{tabular}{l|c|cccccc}
\hline
\textbf{Feature} & \textbf{Mel-spectrogram} & \multicolumn{6}{c}{\textbf{Representation}}                                                                                                                                                                                                                                                                                                                                                                                                                  \\ \hline
\textbf{Type}    & \textbf{Clean}           & \textbf{\begin{tabular}[c]{@{}c@{}}Clean\\ (Layer 0)\end{tabular}} & \textbf{\begin{tabular}[c]{@{}c@{}}Clean\\ (Layer 1)\end{tabular}} & \textbf{\begin{tabular}[c]{@{}c@{}}Clean\\ (Layer 3)\end{tabular}} & \textbf{\begin{tabular}[c]{@{}c@{}}Clean\\ (Layer 5)\end{tabular}} & \textbf{\begin{tabular}[c]{@{}c@{}}Clean\\ (Layer 12)\end{tabular}} & \textbf{\begin{tabular}[c]{@{}c@{}}Clean\\ (Average of all layers)\end{tabular}} \\ \hline
\textbf{SNR}     & 11.08                    & 12.86                                                                 & 12.79                                                                 & 12.80                                                                 & 12.50                                                                 & 13.52                                                                  & 13.13                                                                               \\ \hline
\end{tabular}
\end{table*}

\begin{table*}[!t]
\caption{In the inference stage, the SNR of the synthesized audio using text.}
\label{tab:table2}
\centering
\begin{tabular}{l|cc|cccccc}
\hline
\textbf{Feature} & \multicolumn{2}{c|}{\textbf{Mel-spectrogram}} & \multicolumn{6}{c}{\textbf{Representation}}                                                                                          \\ \hline
\textbf{Type}    & \textbf{Clean}       & \textbf{Enhanced}      & \textbf{\begin{tabular}[c]{@{}c@{}}Enhanced\\ (Layer 0)\end{tabular}} & \textbf{\begin{tabular}[c]{@{}c@{}}Enhanced\\ (Layer 1)\end{tabular}} & \textbf{\begin{tabular}[c]{@{}c@{}}Enhanced\\ (Layer 3)\end{tabular}} & \textbf{\begin{tabular}[c]{@{}c@{}}Enhanced\\ (Layer 5)\end{tabular}} & \textbf{\begin{tabular}[c]{@{}c@{}}Enhanced\\ (Layer 12)\end{tabular}} & \textbf{\begin{tabular}[c]{@{}c@{}}Enhanced\\ (Average of all layers)\end{tabular}} \\ \hline
\textbf{SNR}     & 14.59                & 12.70                  & 13.23                                                                 & 13.07                                                                 & 12.92                                                                 & 13.16                                                                 & 13.34                                                                  & 12.81                                                                               \\ \hline
\end{tabular}
\end{table*}

\subsection{Text-to-representation: FastSpeech 2}
We use FastSpeech 2~\cite{ren2020fastspeech} to learn the mapping from text to representations.
The FastSpeech 2 model mainly consists of phone embedding, encoder, variance adaptor and decoder modules.
The encoder consists of a multi-layer feed-forward transformer, which converts a sequence of phonemes into a sequence of hidden states.
The variance adaptor exploits a multi-layer convolutional network to predict duration, pitch, energy, etc.
The decoder consists of a linear projection layer, which is used to map the network output to the representation.
For more details about FastSpeech 2, please refer to~\cite{ren2019fastspeech, ren2020fastspeech}.
The procedure of training text to representation is shown in Fig.~\ref{fig:figure1}(b).
The noisy speech is first fed into the speech enhancement model to obtain the enhanced speech, which is then fed into the pre-trained model to extract the corresponding representation.
We train the FastSpeech 2 model using paired text and enhanced representations.
It is worth mentioning that both the speech enhancement model and the pre-trained model are publicly available models, and their parameters are not updated.
After training, we input the text and connect the FastSpeech 2 model and the vocoder model to synthesize the speech waveform in the inference stage as shown in Fig.~\ref{fig:figure1}(c).

\section{EXPERIMENTAL setup}
\label{sec:experiment}

\subsection{Model configurations}
For the vocoder, we choose the publicly available multi-speaker clean speech dataset LibriTTS\footnote{https://www.openslr.org/60/}~\cite{zen19_interspeech} train-clean-100 subset in order to ensure that the data for training the vocoder is universal and does not overlap with the dataset for training FastSpeech 2.
In principle, the self-supervised pre-trained model could be any publicly available model, but we use WavLM~\cite{9814838} in this work.
We trained six models, extracting the representations of the 0th (the output of the WavLM feature encoder), 1st, 3rd, 5th, and 12th layers of WavLM and averaging the representations of all layers as input.
All audio sample rates are converted to 24kHz.
We use mel-spectrogram features of clean speech to train a vocoder as the baseline.
For the baseline model, The fast Fourier transform (FFT) size of the extracted mel-spectrogram is set to 1024, the hop size to 240, and the window size to 960. the number of frequency bins of the mel-spectrogram are set to 80, respectively.
For our model, since the frame shift of the representation extracted by WavLM is 20 ms, and the frame shift of mel-spectrogram is 10ms, we set the FFT size to 1024, the hop size to 480, and the window size to 960. The dimension of representation is 768, respectively.
The batch size is set to 16, and a total of 800k steps are trained.
The hyper-parameters $\alpha$ and $\beta$ in (\ref{eq1}) are set to 2 and 45, respectively.

For training FastSpeech 2, we utilize the LJSpeech dataset\footnote{https://keithito.com/LJ-Speech-Dataset/}.
To simulate the noisy environment, we mix the LJSpeech speech data with noise at a signal-to-noise ratio (SNR) of  5 dB as a noisy dataset, where the noise data comes from the Freesound dataset~\cite{2502081}.
To ensure that the speech enhancement model has not seen the LJSpeech dataset, the speech enhancement model\footnote{https://huggingface.co/speechbrain/sepformer-wham16k-enhancement} is publicly available and was trained on other datasets.
The enhanced speech is fed to WavLM to extract the representations of different layers.
We also extract the representations of the 0th, 1st, 3rd, 5th, and 12th layers of WavLM and the representation averaged over all layers.
All audio sample rates are converted to 24kHz.
We train two models using the mel-spectrogram of clean speech and the mel-spectrogram of enhanced speech as the baseline, respectively.
For the baseline model, the FFT size of the extracted mel-spectrogram is set to 1024, the hop size to 240, and the window size to 960. The frequency bins of the mel spectrum are set to 80.
For our model, we set the FFT size to 1024, the hop size to 480, and the window size to 960. The dimension of representation is 768.
The batch size is set to 16, and a total of 900k steps are trained, respectively.

\subsection{Evaluation metrics}
For different models, we respectively generated 256 utterances from the test set.
To measure the level of purity of the speech generated using mel-spectrogram and representation features, we use the SNR.
We test the SNR of the waveform generated by the vocoder and the SNR of the waveform generated by the whole TTS model.
In addition, for the objective evaluation metric, we evaluate the Mean Opinion Score - Listening Quality Objective (MOS-LQO) using the VISQOL\footnote{https://github.com/google/visqol}~\cite{9123150} tool, where the MOS-LQO value ranges from 1 to 4.75, i.e., a higher value indicates a better speech quality.
As the subjective evaluation, we test the mean opinion score (MOS) to evaluate the naturalness and robustness of the speech, which ranges from 1 to 5 (the higher, the better).

\section{EXPERIMENTAL result}
\subsection{Evaluation  of the vocoder}
We utilize the mel-spectrogram based vocoder and FastSpeech 2 as baseline models.
In order to test the denoising performance of the vocoder model, we feed the mel-spectrogram features of 5dB noisy speech into the trained vocoder using the train-clean-100 subset of LibriTTS, and calculate the SNR of the generated speech. The results are shown in Table~\ref{tab:table1}.
Similarly, we input the representation of 5dB noisy speech into the vocoder model, and calculate the SNR of the speech generated by the vocoder, and the results are shown in Table~\ref{tab:table1}.
The synthesized speech can be obtained from https://zqs01.github.io/rep2wav/.
It is clear that the vocoder trained using representation obtains a higher SNR than using mel-spectrogram, which implies that representation has a better noise robustness.
In addition, we find that the vocoder trained with the representation of layer 12 has the best noise robustness with an SNR of 13.52 for the generated speech.
However, from the viewpoint of auditory perception,  although the representation of layer 12 has a stronger ability to suppress noise, it also suppresses more speaker's information, i.e., the partial loss of speaker information leads to a slight change in the speaker in the synthesized speech, as shown in Table~\ref{tab:table5}.
For this, we train the vocoder using a representation averaged over all layers, resulting in an SNR of 13.13,  which can well balance the noise robustness and the preservation of speaker information.
In addition, we connect text-to-representation and representation-to-wav modules trained using the enhanced speech and then test the SNR of the synthesized speech. The experimental results are shown in Table~\ref{tab:table2}, which similarly demonstrate that the noise robustness of the representation of layer 12 is the best.

\begin{table}[!t]
\caption{In the inference stage, the MOS-LQO of the synthesized audio using text.}
\label{tab:table3}
\centering
\scalebox{0.95}{
\begin{tabular}{l|c|c}
\hline
\textbf{Feature}                 & \textbf{Type}                    & \textbf{MOS-LQO} \\ \hline
\multirow{2}{*}{Mel-spectrogram} & Clean                            & 3.78             \\
                                 & Enhanced                         & 2.58             \\ \hline
\multirow{6}{*}{Representation}  & Enhanced (Layer 0)               & 2.97             \\
                                 & Enhanced (Layer 1)               & 3.03             \\
                                 & Enhanced (Layer 3)               & 3.13             \\
                                 & Enhanced (Layer 5)               & 3.21             \\
                                 & Enhanced (Layer 12)              & 3.05             \\
                                 & Enhanced (Average of all layers) & 3.32             \\ \hline
\end{tabular}}
\end{table}

\begin{table}[]
\caption{In the inference stage, the MOS of the synthesized audio using text.}
\label{tab:table4}
\centering
\begin{tabular}{l|c|c}
\hline
\textbf{Feature}                 & \textbf{Type}                    & \textbf{MOS} \\ \hline
-                                & Ground Truth                     & 4.16         \\ \hline
\multirow{2}{*}{Mel-spectrogram} & Clean                            & 3.84         \\
                                 & Enhanced                         & 2.65         \\ \hline
\multirow{6}{*}{Representation}  & Enhanced (Layer 0)               & 3.38         \\
                                 & Enhanced (Layer 1)               & 3.74         \\
                                 & Enhanced (Layer 3)               & 3.86         \\
                                 & Enhanced (Layer 5)               & 3.89         \\
                                 & Enhanced (Layer 12)              & 3.73         \\
                                 & Enhanced (Average of all layers) & 3.80         \\ \hline
\end{tabular}
\end{table}

\subsection{Subjective and objective evaluations}

The subjective performance of the synthesized speech in terms of  MOS-LQO is shown in Table~\ref{tab:table3}.
The baseline model trained with clean speech achieves a MOS-LQO of 3.78 and the baseline model trained with enhanced speech obtains a MOS-LQO of 2.58. This is due to the fact that the TTS models trained using the enhanced speech often contain noise (i.e., speech distortions), leading to low MOS-LQO values.
The models trained with the representation generally outperform the baseline model, and the one trained with the representation averaged over all layers achieved the best performance with a MOS-LQO of 3.32.
Compared to the baseline model, the noise component in the synthesized speech is significantly reduced when the representation is used.
The quality of the synthesized audio is similar using the representation of middle layers, e.g., the 3rd, and 5th layers.
In addition, models trained using representations averaged over all layers can balance the noise robustness and speaker information well.

We show the MOS scores of the speech generated by the different models in Table~\ref{tab:table4}.
It is clear that the MOS scores of the synthesized speech from the TTS model trained with representation are higher than that of the synthesized speech from the TTS model trained with mel-spectrogram.
The qualities of the synthesized speech using the representation of layers 3 and 5 are similar and comparable to that of using the average of all layers.

\begin{table}[!t]
\caption{Speaker similarity of synthesized speech using different models.}
\label{tab:table5}
\centering
\scalebox{0.95}{
\begin{tabular}{l|c|c}
\hline
\textbf{Feature}                 & \textbf{Type}                    & \textbf{\begin{tabular}[c]{@{}c@{}}Speaker\\ similarity\end{tabular}} \\ \hline
\multirow{2}{*}{Mel-spectrogram} & Clean                            & 0.8363                                                                      \\
                                 & Enhanced                         &   0.5550                                                                    \\ \hline
\multirow{6}{*}{Representation}  & Enhanced (Layer 0)               & 0.5593                                                                \\
                                 & Enhanced (Layer 1)               & 0.5445                                                                \\
                                 & Enhanced (Layer 3)               & 0.5230                                                                \\
                                 & Enhanced (Layer 5)               & 0.5040                                                                \\
                                 & Enhanced (Layer 12)              & 0.2466                                                                \\
                                 & Enhanced (Average of all layers) & 0.4687                                                                \\ \hline
\end{tabular}}
\end{table}

\begin{figure}[!t]
\centering
\includegraphics[width=0.45\textwidth]{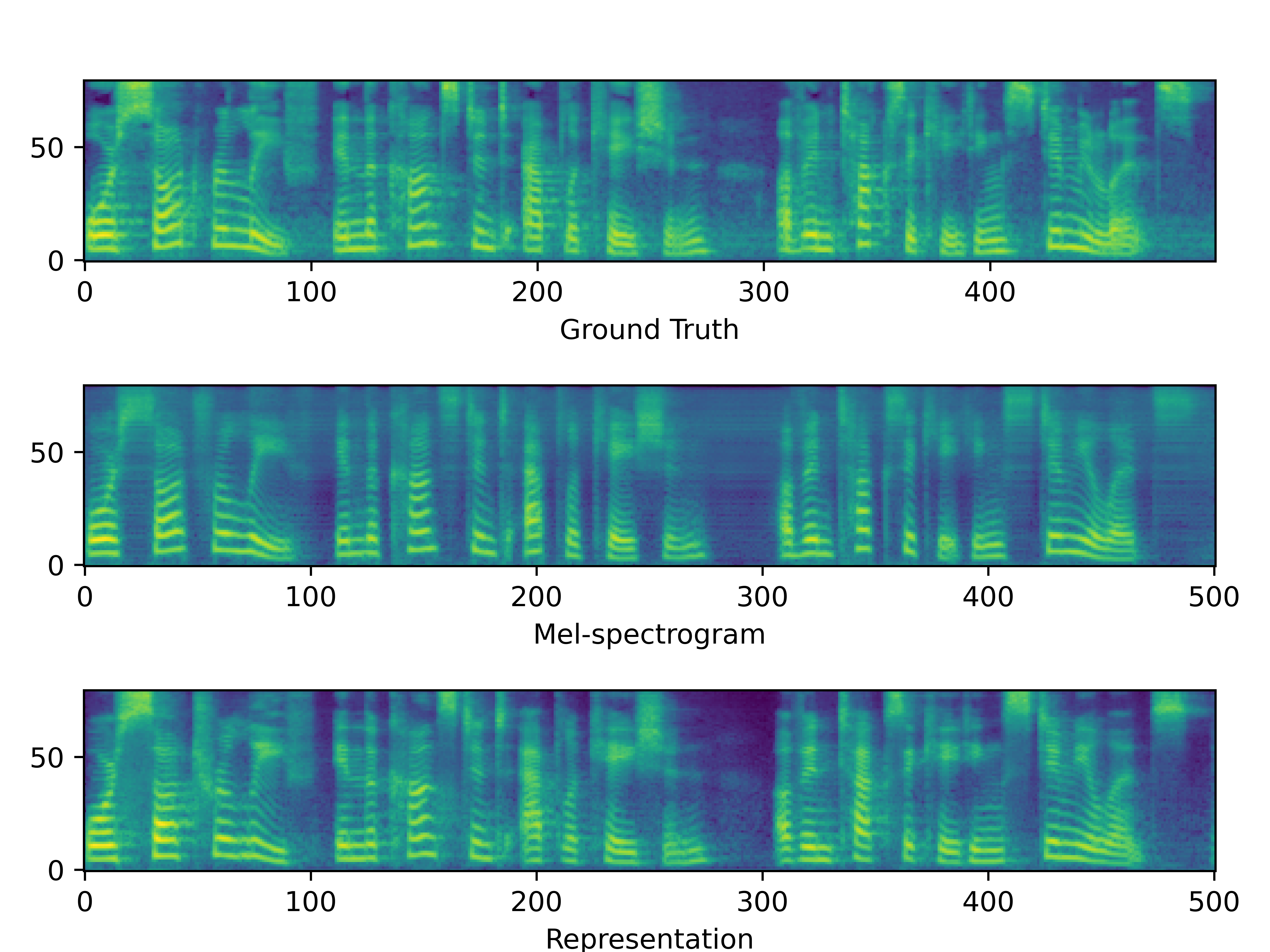}
\caption{The mel-spectrogram of the generated speech using the TTS model trained with mel-spectrogram or representation feature.}
\label{fig:figure2}
\end{figure}

\subsection{Visualization of the generated speech}
Finally, we visualize the speech synthesized by the model trained with mel-spectrogram and the model trained with representation in Fig.~\ref{fig:figure2}.
The top shows the mel-spectrogram of the clean speech, the middle is the mel-spectrogram obtained by the TTS model trained using the mel-spectrogram of the enhanced speech, and the bottom shows the mel-spectrogram for speech generated by the TTS model trained using the representation of the enhanced speech. We can see that the speech generated by the model trained with representations has fewer noise components and is more similar to clean speech.

\section{CONCLUSION}
\label{sec:conclusion}
In this paper, we investigated the representation-based noise robust text-to-speech model.
By constructing representation-to-waveform vocoder and text-to-representation FastSpeech 2 models, we found that the representation-based TTS model has better noise robustness than the conventional mel-spectrogram-based TTS model.
In addition, high-level representation can suppress the noise component better, but also cause a slight loss of speaker information.
Averaging the representations of all layers provides a good balance between the noise robustness and speaker information.

	
\bibliographystyle{IEEEbib}
\bibliography{strings,refs}
	
\end{document}